\documentclass[superscriptaddress,twocolumn,secnumarabic,
amssymb,amsmath,nobibnotes,aps,prd,nofootinbib]{aastex62}
\graphicspath{{./}{figures/}}

\usepackage[utf8]{inputenc}
\usepackage{graphicx}
\usepackage{dcolumn}
\usepackage{bm}
\usepackage{ulem} 
\usepackage{hyperref}
\usepackage{multirow,enumitem}
\usepackage{amsmath}
\usepackage{color}
\usepackage{xcolor}
\usepackage{multirow}

\begin{document}
\title{Measuring the speed of light with updated Hubble diagram of high-redshift standard candles}

\author{Yuting Liu}
\affiliation{Institute for Frontiers in Astronomy and Astrophysics, Beijing Normal University, Beijing 102206, China;}
\affiliation{Department of Astronomy, Beijing Normal University, Beijing 100875, China;}
\affiliation{Department of Physics, University of Tokyo, Tokyo, 113-0033, Japan;}

\author{Shuo Cao}
\altaffiliation{caoshuo@bnu.edu.cn}
\affiliation{Institute for Frontiers in Astronomy and Astrophysics, Beijing Normal University, Beijing 102206, China;}
\affiliation{Department of Astronomy, Beijing Normal University, Beijing 100875, China;}

\author{Marek Biesiada}
\affiliation{National Centre for Nuclear Research, Pasteura 7, Warsaw, 02-093, Poland}

\author{Yujie Lian}
\affiliation{Institute for Frontiers in Astronomy and Astrophysics, Beijing Normal University, Beijing 102206, China;}
\affiliation{Department of Astronomy, Beijing Normal University, Beijing 100875, China;}

\author{Xiaolin Liu}
\affiliation{Institute for Frontiers in Astronomy and Astrophysics, Beijing Normal University, Beijing 102206, China;}
\affiliation{Department of Astronomy, Beijing Normal University, Beijing 100875, China;}

\author{Yilong Zhang}
\affiliation{Institute for Frontiers in Astronomy and Astrophysics, Beijing Normal University, Beijing 102206, China;}
\affiliation{Department of Astronomy, Beijing Normal University, Beijing 100875, China;}

\begin{abstract}
The possible time variation of the fundamental constants of nature has been an active subject of research in modern physics. In this paper, we propose a new method to investigate such possible time variation of the speed of light $c$ using the updated Hubble diagram of high-redshift standard candles including Type Ia Supernovae (SNe Ia) and high-redshift quasars (based on UV-X relation). Our findings show that the SNe Ia Pantheon sample, combined with currently available sample of cosmic chronometers, would produce robust constraints on the speed of light at the level of $c/c_0=1.03\pm0.03$. For the Hubble diagram of UV+X ray quasars acting as a new type of standard candles, we obtain $c/c_0=1.19\pm0.07$. Therefore, our results confirm that there is no strong evidence for the deviation from the constant speed of light up to $z\sim 2$. Moreover, we discuss how our technique might be improved at much higher redshifts ($z\sim5$), focusing on future measurements of the acceleration parameter $X(z)$ with gravitational waves (GWs) from binary neutron star mergers. In particular, in the framework of the second-generation space-based GW detector, DECi-hertz Interferometer Gravitational-wave Observatory (DECIGO), the speed of light is expected to be constrained with the precision of $\Delta{c}/c=10^{-3}$.

\end{abstract}

\section{Introduction}\label{sec:intro}

As a test of fundamental physics, probing the space-time variation of fundamental constants of Nature (the fine-structure constant $\alpha$, the speed of light $c$, the proportionality constant $G$, the Planck constant $\hbar$) has been undertaken in the past decades, following the pioneering work of  \citet{Dirac(1934)}. In particular, experiments on Earth and solar system have been designed and carried out for centuries to measure the speed of light $c$, with extreme precision in their measurements. However, the assumption that the speed of light $c$ is constant across all universe at every time and distance scale is a far reaching extrapolation of our knowledge, grounded on a fairly limited space-time region. Even Einstein himself had already considered the theory of the dynamic speed of light \citep{Einstein1907}. The speed of light may, in principle change over time during the evolution of the universe. Such an idea is referred to as the variable speed of light (VSL) theory \citep{Albrecht1999,Magueijo1999}, which has attracted considerable attention some time ago because it can provide a new way of solution of classical cosmological problems such as the initial singularity, horizon and flatness problems, without relying on inflationary scenarios
\citep{Albrecht1999,Barrow1999,Magueijo1999,Bassett2000}. Similar ideas were independently formulated and strongly supported by \citet{Moffat2002}. Later studies revealed that variable $c$ theories are able to explain the scale-invariant spectrum of Gaussian fluctuations in cosmic microwave background (CMB) data \citep{Magueijo2003}. On the other hand, some authors claimed that dimensional constants like $c$ are merely human constructions, contrary to their dimensionless combinations like the fine structure constant $\alpha$ \citep{Duff2002}. Hence, their temporal changes have no operational significance. Yet, the variability of the speed of light is still a controversial issue.

At present, the measurements of the speed of light on the Earth have reached a very high accuracy and support its being a fundamental constant. Still, cosmological tests of $c$ variations are much more scarce and less precise. Fortunately, a variety of high-quality observational data obtained from extra-galactic surveys are becoming available to test the basic laws in the more distant universe \citep{Salzano2016,Balcerzak2017,Salzano2017,Salzano2017a,Salzano2017b}. Based on a flat FLRW universe, a simple relationship $c(z_M)=D_A(z_M)H(z_M)$ between the speed of light $c$, the angular diameter distance $D_A(z)$ and the Hubble parameter $H(z)$ was proposed \citep{Salzano2015}, at the redshift $z_M$ corresponding to the maximum of $D_A(z)$ function. \citet{Cai2016} developed a new and more general approach to test the speed of light in the larger redshift range, using the luminosity distance $D_L(z)$ from type Ia supernovae sample. Subsequently, \citet{Cao2017} obtained the measurement of the speed light using the maximum redshift $z_M=1.70$ obtained from the intermediate-luminosity compact radio quasars acting as standard rulers. The result was absolutely consistent with the value of $c_0$ measured on Earth. \citet{Salzano2017b} also investigated the invariance of the speed of light at different redshifts not only at the specific redshit $z_M$. More recently, \citet{Cao2020,Liu2021} studied the possibility of using strong gravitational lensing to test the invariance of the speed of light. Some other observational tests on the invariance of the speed of light in cosmology have also been performed in \citet{Qi2014,Cao2018,Wang2019}. Of course, we still need to develop other methods to measure the speed of light in the distant universe, which is an almost unexplored domain.

In this context of the discussion presented above, we will focus our attention on an original model-independent technique, which delivers estimates of the speed of light at different redshifts in the distant universe, combining current observations of the standard candles (type \uppercase\expandafter{\romannumeral1}a supernovae, the X-UV relation of quasars) and standard clock data (Hubble parameters $H(z)$ inferred from cosmic chronometers). Since the cosmic chronometers data are currently available up to the redshift $z\sim 2$, we study another possibility, which is to use the future space gravitational wave detector DECIGO allowing to determine the so called acceleration parameter $X(z)$, related to $H(z)$ up to the redshift $z=5$. In this way one might be able to measure the speed of light with considerable accuracy up to the higher redshifts and enrich the estimates of the speed of light in unexplored domain. We would like to emphasize that our technique is completely independent of the details of the cosmological model, and the only assumption we made is the flat FLRW metric. This paper is organized as follows. In Sect. 2, we introduce the methodology and observational data used in this work. Simulations are presented in Sect. 3. Furthermore, our results and general conclusions are summarized in Sect. 4.

\section{Methodology and Data}\label{sec:data}

It is well known that under the assumption of homogeneity and isotropy in the large scale, the geometry of the universe can be described by the FLRW metric
\begin{equation}
ds^2 = - c^2 dt^2 + a^2(t)\left[\frac{dr^2}{1-Kr^2}+r^2(d\theta^2+sin^2\theta d\phi^2)\right],
\end{equation}
where $t$ is the cosmic time and ($r$,$\theta$,$\phi$) are the comoving spatial coordinates. By virtue of the Einstein equations, the scale factor $a(t)$ as the only gravitational degree of freedom can be determined by the matter and energy content of the Universe. The curvature parameter $\Omega_k$ is related to the dimensionless curvature $K$ as $\Omega_k=-cK/(a_0H_0)^2$, where $H_0$ denotes the Hubble constant and $K=-1,0,+1$ corresponds to open, flat and closed Universe, respectively. Considering a flat universe in such metric, the luminosity distance $D_L(z)$ can be expressed as
\begin{equation}\label{eq2}
D_L(z) = c(1+z) \int_{0}^{z}\frac{dz'}{H(z')},
\end{equation}
where $H(z)$ denotes the expansion rate of the Universe at redshift $z$.

Generally, one can introduce the time variable $c$ in the metric or in the Friedmann equations, and assume the speed of light is a function of cosmic time or redshift $z$ in a homogenous and isotropic universe \citep{Qi2014}. In this paper, we use $c$ to quantify the speed of light related to the baseline from redshift $z$ to the Earth (at redshift $z=0$). Any evidence in favour of violating the constancy of $c$ at different redshifts will have a profound impact on our understanding of nature. Differentiating Eq.(2) with respect to the redshfit $z$, one can obtain
\begin{equation}\label{eq3}
D_L^{'}(z) = \ c\int_{0}^{z}\frac{dz'}{H(z')}+c(1+z)\frac{1}{H(z)},
\end{equation}
From this equation one is able to express the speed of light $c$ in terms of the the Hubble parameter $H(z)$, luminosity distance $D_L(z)$ and its derivative $D_L^{'}(z)=dD_L(z)/dz$:
\begin{equation}\label{eq4}
c=\frac{D_L^{'}(z)H(z)}{1+z}-\frac{D_L(z)H(z)}{(1+z)^{2}}.
\end{equation}
Thus, we would be able to determine the speed of light at any single redshift, provided we have all aforementioned ingredients. More importantly, one does not need to assume any particular cosmological model besides flat FLRW metric. In this paper, we focused on an empirical fit to the luminosity distance measurements, based on a third-order logarithmic polynomial of \citep{Risaliti2018,Liu2020,Zheng2021}
\begin{equation}\label{eq5}
D_L(z)=\ln(10) \frac{c_0}{H_0}(x+ax^2+bx^3),
\end{equation}
where $c_0 = 299792458 \; m s^{-1}$ denotes the laboratory value of the speed of light, $x = \log(1+z)$, and $a$ and $b$ are the free parameters. The logarithmic parameterization has the advantage of faster convergence at high redshifts ($z>1$).
Combining Eq.(4) and Eq.(5), one can derive
\begin{equation}\label{eq6v2}
\eta \equiv \frac{c}{c_0}= \frac{y}{(1+z)^2} \left[ 1 + (2a-1)x + (3b-a)x^2 - bx^3 \right],
\end{equation}
where $y = ln(10) H(z)/H_0$. The constancy of the speed of light means $\eta = 1.$
Any deviation of $\eta$ from 1, at some redshift $z^{*}$ will indicate that $c(z^{*})$ is different from $c_0$. Therefore, we can test the invariance of the speed of light at any redshift based on Eq.(6).

In this paper we determine the parameters $a$ and $b$, as best fits to the data regarding standard candles: SNe Ia and quasars with a calibrated UV-X relation, as will be presented below. We use for this purpose the Markov Chain Monte Carlo (MCMC) method implemented in the Python module~\footnote{https://pypi.python.org/pypi/emcee} {\tt{emcee}}  introduced by \citet{Foreman-Mackey2013}.
The next ingredient we need is the expansion rate $H(z)$ at redshift $z$ and it will be extracted from cosmic chronometers. We will also consider the utility of the $H(z)$ related expansion parameter, which can be obtained with the future GW space-borne detector DECIGO. Throughout this work, we take the prior on the Hubble constant $H_0=67.4\pm0.5$ km/s/Mpc from the latest \textit{Planck} CMB observations \citep{Aghanim2020}.


\subsection{The Pantheon Sample of Type  \uppercase\expandafter{\romannumeral1}a Supernovae}\label{subsec:Supernovae}

Type \uppercase\expandafter{\romannumeral1}a Supernovae  (SNe \uppercase\expandafter{\romannumeral1}a) regarded as the standard candles for their  standardizable luminosity have been used to discover the accelerating expansion of the Universe \citep{Riess1998,Perlmutter1999} and are widely used as cosmological probes.
Many supernovae surveys have been focused on detecting supernovae within a considerable range of redshifts over the past two decades, including low-redshift ($0.01<z<0.1$) surveys, e.g. CfA1-CfA4, CSP and LOSS \citep{Riess1999,Jha2006,Stritzinger2011} and  four main surveys probing the $z>0.1$ redshift range like ESSENCE, SNLS, SDSS and PS1 \citep{Miknaitis2007,Conley2011,Frieman2008,Scolnic2014}.
Morever, SCP, GOODS and CANDELS/CLASH surveys released the high-z ($z>1.0$) data  \citep{Suzuki2012,Riess2004,Riess2007,Rodney2014}. These surveys extended the Hubble diagram to $z = 2.26$.
\begin{figure}
\begin{center}
\includegraphics[width=0.95\linewidth]{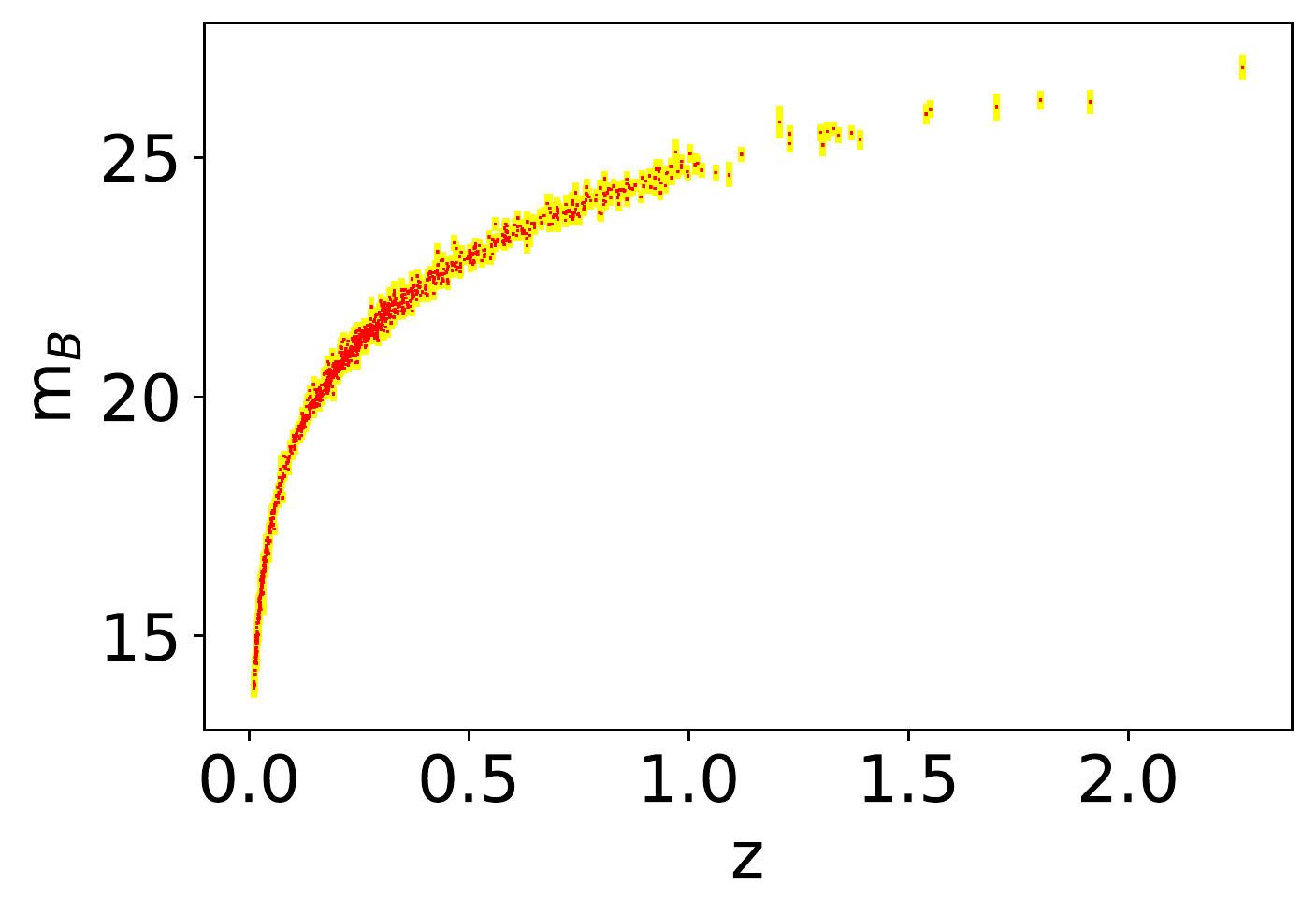}
\end{center}
\caption{Scatter plot of the type \uppercase\expandafter{\romannumeral1}a supernovae from the Pantheon sample.
The red points denote the apparent B-band magnitude of the type \uppercase\expandafter{\romannumeral1}a supernovae, with its 1$\sigma$ confidence level (yellow bars).}
\end{figure}

More recently, \citet{Scolnic2018} combined the subset of 279 Pan-STARRS1(PS1) ($0.03 < z < 0.68$) supernovae \citep{Rest2014,Scolnic2014} with the data from SDSS, SNLS, various low-z and HST samples to form the largest combined sample of SNe Ia consisting of a total of 1048 SNe Ia ranging from $0.01<z<2.3$, which is known as  the ``Pantheon Sample''. Systematic uncertainties in the Pantheon sample have been reduced by the improvements of the PS1 SNe photometry, astrometry and calibration.
Generally speaking, the light curve of Type Ia SN is characterized by 3 or 4 nuisance parameters, and its use involves their optimization together with the unknown parameters of the cosmological model. Fortunately, \citet{Kessler2017} proposed a new method (i.e., BEAMS with Bias Corrections (BBC)) to retrieve the nuisance parameters in the Tripp formula \citep{Tripp1998}
\begin{equation}
\mu=m_B - M + \alpha {x_1}  -\beta^{*} {C} + \Delta {M} + \Delta {B},
\end{equation}
where $\mu$ is the distance modulus, $m_B$ is the apparent B-band magnitude, $C$ is the color, $x_1$ is the light-curve shape parameter, $\Delta {M}$ is a distance correction based on the host-galaxy mass of the SNe and $\Delta {B}$ is a distance correction based on predicted biases from simulations. Furthermore, $\alpha$ is the coefficient of the relation between luminosity and stretch, $\beta^{*}$ is the coefficient of the relation between luminosity and color and $M$ is the absolute B-band magnitude of a fiducial SN Ia with $x_1= 0$ and $C = 0$. The Pantheon Sample is relatively clean and the obvious advantage of using it is its richness and depth in redshift as compared with the previous data sets such as Union2.1 or JLA \citep{Betoule2014}. We show the scatter plot of the 1048 Pantheon Sample in Figure 1.

As already mentioned, we aim to employ Eq.(\ref{eq6v2}) for the assessment of the invariance of the speed of light. For this purpose we need to use the  high-redshift Pantheon Sample to fit the free parameters $a$ and $b$ in the phenomenological representation of the luminosity distance, by minimizing the $\chi^{2}$ objective function
\begin{equation}
\chi^2=\sum \limits_{i=1}^{1048}{\frac{[D_{L,SNe}(z_i)-D_L(z_i;a,b)]^2}{\sigma_{D_{L,SNe}}(z_i)^2}},
\end{equation}
where  $D_{L,SNe}$ (in $Mpc$) is calculated from distance modulus using a well-known relation:
\begin{equation}
D_{L,SNe}=10^{0.2(m-M) - 5},
\end{equation}
and the uncertainty of luminosity distance from SNe is given by
\begin{equation}
\sigma_{D_{L,SNe}}=(ln10/5)D_{L,SNe}\sigma_{*i},
\end{equation}
where $\sigma_{*i}$ is total uncertainty of the apparent magnitude.
We treat absolute magnitude $M$ as a free parameter fitted together with the free parameter $(a,b)$ characterizing the luminosity distance. Such methodology, which generated the best fitted values with 68\% C.L  of $a = 3.15\pm0.12$, $b = 3.27\pm0.41$, and $M=-19.45\pm0.01$.

\subsection{The Hubble diagram of High-redshift Quasar}\label{subsec:QSO}
Quasars as the brightest sources in the Universe that can be observed up to redshift $z\sim 8.0$, have long been attempted to be used as potential standard candle candidates for extending the distance range as compared with supernovae \citep{Mortsell2011,Baados2018}.
The standard candle suitable for cosmological research has to have two basic properties: one -- it has to have a standard (or standardizable) intrinsic luminosity; second -- it should be easy to observe in a wide redshift range. Quasars are one of the best candidates satisfying the latter, but they do not clearly display the former property. Luminosity of the quasars emission regions spans several orders of magnitude, hence at the first glance they
appear to have little chance of becoming the standard candle. However, with a large enough sample of quasars one may attempt to discover correlations between luminosities at different spectral bands, try to select a sample with not too large dispersion and use it as a cosmological tool.

\begin{figure}
\begin{center}
\includegraphics[width=0.95\linewidth]{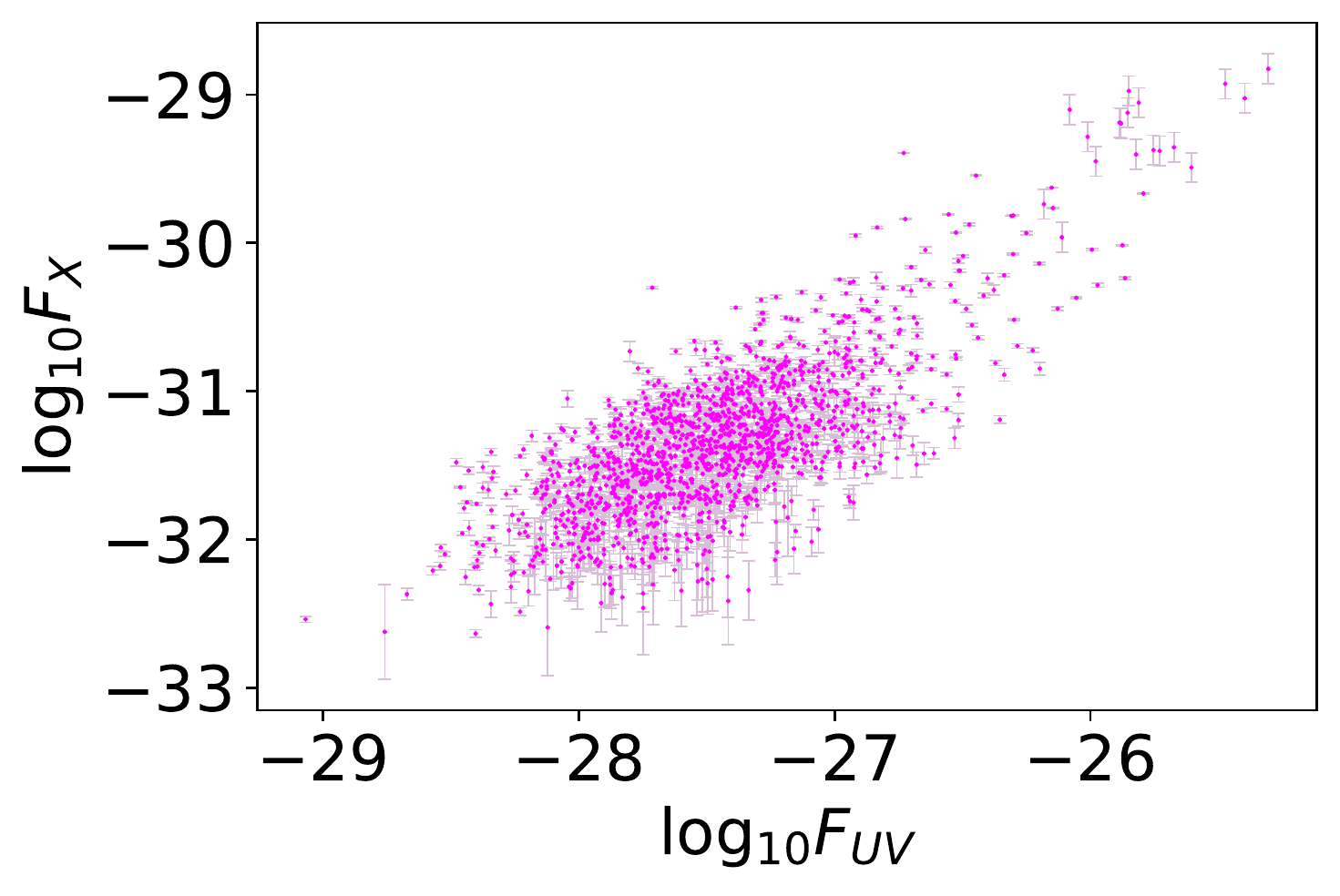}
\end{center}
\caption{Scatter plot of the UV+X ray quasar sample. The pink points denote the logarithm of the flux in X-ray and UV. The gray bars represent $F_X$ 1$\sigma$ uncertainty level. The uncertainty of $F_{UV}$ measurements are negligible.}
\end{figure}

Along this line of reasoning, the non-linear relationship between the quasars ultraviolet and X-ray luminosity ($L_{UV}-L_{X}$) has been proposed as such a tool. According to the unified picture of active galactic nuclei (AGN), quasars in particular, it is generally believed that ultraviolet photons are emitted from the accretion disk, while X-rays originate from inverse Compton scattering of UV photons passing through the hot corona. Therefore one is encouraged to seek for the non-linear relation between the X-ray and UV luminosities of quasars, parameterized as a linear one in logarithmic variables:
\begin{equation}
\log_{10}(L_X)=\gamma \log_{10}(L_{UV})+\beta,
\end{equation}
where $L_X$ and $L_{UV}$ correspond to the rest-frame monochromatic luminosity at $2 \; keV$ and $2500\; \text{\AA}$, the slope $\gamma$ and the intercept $\beta$ are free parameters which can be calibrated with other cosmological probes.
On the other hand, the observable quantity is flux, not luminosity, hence the non-linear relation between the X-ray and UV fluxes can be written as
\begin{equation} \label{FX-FUV}
\log_{10}(F_X)=\gamma\log_{10}(F_{UV})+2(\gamma-1)\log_{10}(D_L)+\beta',
\end{equation}
where $F_X$ and $F_{UV}$ correspond to the  X-ray flux and UV flux, respectively $\beta^{'}=(\gamma-1)\log_{10}(4\pi)+\beta$ is the intercept of this new relation and $D_L$ is luminosity distance. Therefore, the correlation between the observed UV and X radiation fluxes can be used to obtain the luminosity distance.

\citet{Risaliti2015} collected a sample of 1138 quasars from large surveys such as COSMOS, SDSS, XMM and for the first time used the nonlinear relationship between the UV and X-ray radiation flux to estimate cosmological distances in the high redshift range. However, the high dispersion (0.35 -- 0.40 dex) revealed in the quasar data was a major obstacle for cosmological applications.
In the subsequent series of studies \citep{Lusso2016,Bisogni2017}, the dispersion in quasars data set has been significantly reduced.

Recently, \citet{Risaliti2018} have taken a key step in the study of the non-linear $L_{UV}-L_{X}$ relationship in quasars to construct the Hubble diagram, covering the redshift range from $z=0.04$ to $z=5.1$. The final sample of 1,598 sources they used to build the Hubble diagram has been obtained from a parent sample of 7,237 quasars. Selection criteria used to derive the final sample included consideration of X-ray absorption, interstellar reddening effects, pollution of UV observations, Eddington deviation, etc. The final sample is built by merging the following groups of quasars: 791 sources from the SDSS-DR7 sample, 612 from the SDSS-DR12, 102 from XMM-COSMOS, 18 from the low-redshift sample (Swift), 19 from Chandra-Champ, 38 from the high-z ($z>4$) sample, and 18 quasars from the new z$\sim$3 sample (XMM-Newton Very Large Program). After this selection, the intrinsic dispersion was reduced to 0.23 dex. Many international project teams have been using this updated Hubble diagram of high-redshift standard candles in cosmological research  \citep{Liao2019,Melia2019,Liu2020a,Yang2020,Geng2020,Zheng2021,Borislavov2021,Lian2021,Zhao2021,Li2021}. We show the scatter plot of the 1598 quasar sources in Figure 2.

As previously, in the case of Pantheon SN Ia sample, we use the updated UV-X quasar data to fit the parameters $a$ and $b$ characterizing the phenomenological form of the luminosity distance. They are obtained by minimizing the following objective function
\begin{equation}
\chi^2=\sum \limits_{i=1}^{1598}{\frac{[D_{L,QSO}(z_i)-D_L(z_i;a,b)]^2}{\sigma_{D_{L,QSO}}(z_i)^2}},
\end{equation}
where $D_{L,QSO}(z)$ obtained from Eq.(\ref{FX-FUV}) can be expressed as
\begin{equation}
D_{L,QSO}=10^{\frac{1}{2-2\gamma} [\gamma\log(F_{UV})-\log(F_X) + \beta' ]}
\end{equation}
The uncertainty of the luminosity distance obtained from QSOs is given in terms of the $(F_X)_i$ measurement uncertainty $\sigma_i$ and the global intrinsic dispersion $\delta$:
\begin{equation}
\sigma_{D_{L,QSO}}=\frac{ln(10)\; D_{L,QSO}}{2-2\gamma}\sqrt{\sigma^{2}_{i}+\delta^2}
\end{equation}
The uncertainty of $F_{UV}$ is negligible comparing to $\sigma_i$ and $\delta$, and is therefore ignored in this paper \citep{Risaliti2015}. During the fit we also treat $(\gamma,\beta,\delta)$  as free parameters fitted together with the parameters $(a,b)$ characterizing the luminosity distance. In this case, we obtain the best fitted values of free parameters with $1\sigma$
error, respectively, $a = 5.33^{+1.75}_{-1.38}$, $b =-0.16^{+1.30}_{-1.50}$,  $\gamma = 0.62\pm0.01$, $\beta = 7.83\pm0.29$ and $\delta = 0.23\pm0.003$.

\subsection{Cosmic Chronometers $H(z)$}\label{subsec:H(z)}

The Hubble parameter $H(z) = \dot{a}/a$ characterizes the expansion rate of the universe at given redshift and has recently been
widely used in cosmological research.
A model-independent procedure of measuring the expansion rate of the universe has been proposed by \citet{Jimenez2002},
using the differential age of passively evolving galaxies. This is known as cosmic chronometer approach and is based on the relation
\begin{equation}\label{eq13}
H(z)=-\frac{1}{1+z}\frac{dz}{dt}.
\end{equation}
From the measurements of the age difference $\Delta t$ between two passively evolving galaxies separated by a small redshift interval $\Delta z$, one can approximate the derivative $dz/dt$ by the ratio $\Delta z/\Delta t$.
In order to accurately calculate the Hubble parameter $H(z)$, the average age of stars in each galaxy should be much larger than the age difference between the galaxies. Galaxy pairs selected as cosmic chronometers, besides being close in redshift, should meet the following two conditions: similar metal abundance and low star formation rate. Therefore, it is necessary to select those passively evolved galaxies with reddish spectra dominated by the old population.

Currently popular approach is to determine the age of passively evolving galaxies from the spectral feature known as $4000\;\text{\AA}$ break. Denoted as $D4000$, the feature is defined as
the ratio between the continuum flux densities in a red band and a blue band around $4000\;\text{\AA}$. It originates from a series of metal absorption features, and is known to correlate with the stellar metallicity and age of the stellar population \citet{Moresco2016}. Historically, in short, \citet{Simon2005} analyzed Gemini Deep Survey (GDDS) and archival data to obtain 8 $H(z)$ data points, which they used to constrain the dark energy. \citet{Stern2010} improved previous expansion history measurements of \citet{Simon2005} by the high-quality spectra with the Keck-LRIS spectrograph of red-envelope galaxies in 24 galaxy clusters in the redshift range $0.2 < z < 1.0$ from the SPICES and VVDS surveys. \citet{Chuang2012} measured the $H(z)$ at $z=0.35$ based on the Sloan Digital Sky Survey Data Release 7 data.
\citet{Moresco2012} obtained 8 new $H(z)$ data points in the redshift range $0 < z < 1.75$ and expanded the sample size to 20.

\begin{figure}
\begin{center}
\includegraphics[width=0.95\linewidth]{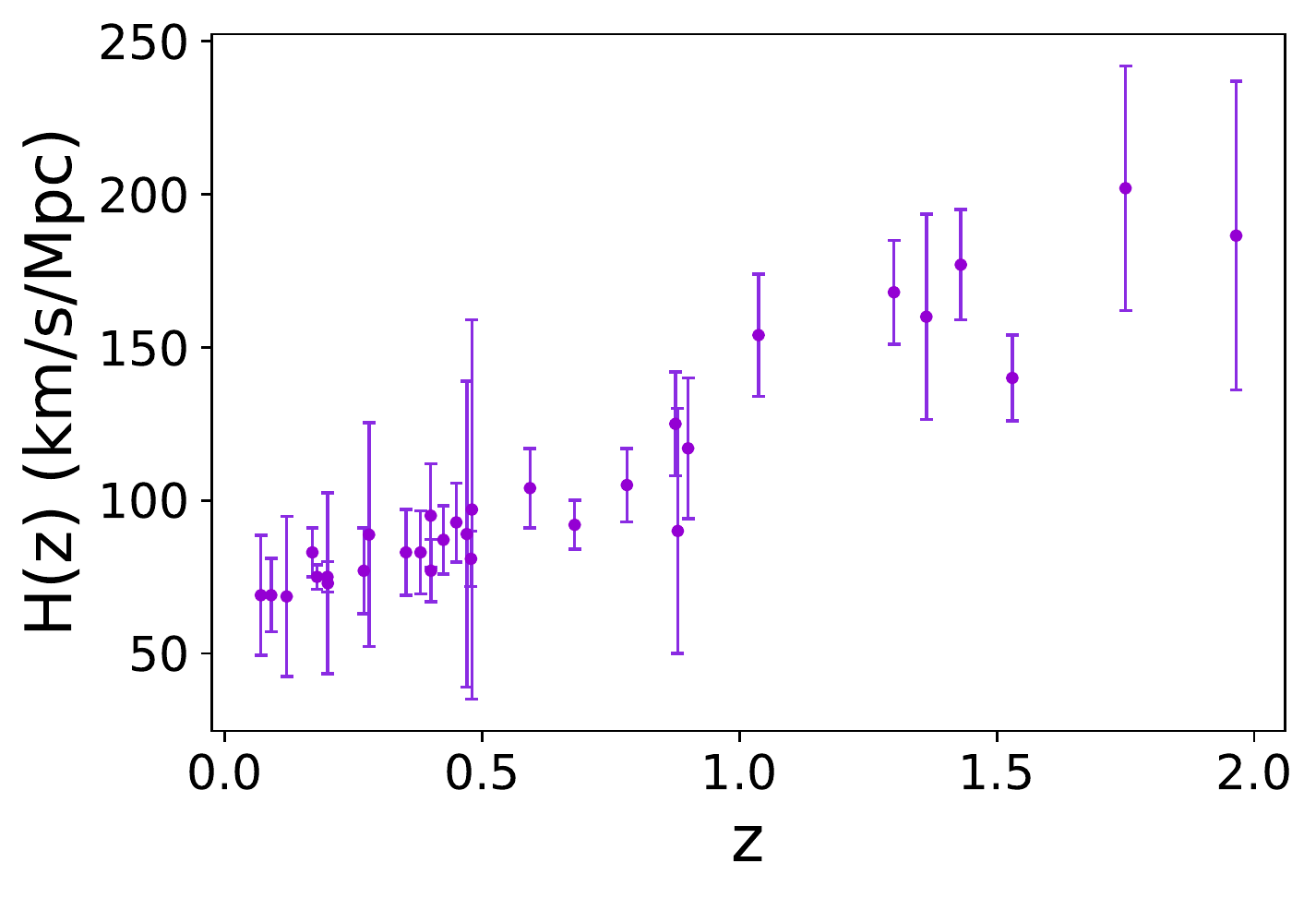}
\end{center}
\caption{Scatter plot of the 31 cosmic chronometers sample. Bars
denote the Hubble parameters $H(z)$, with their 1$\sigma$ uncertainties.}
\end{figure}

Then, \citet{Zhang2014} obtained 4 new observational $H(z)$ data from 17,832 luminous red galaxies covering redshift $0 < z < 0.4$ in the Sloan Digital Sky Survey DR7. More recently, 2 new measurements of the Hubble parameter $H(z)$ were presented by \citet{Moresco2015} using the cosmic chronometer method up to $z\sim2$. \citet{Moresco2016} exploited the unprecedented statistics provided by the Baryon Oscillation Spectroscopic Survey (BOSS) Data Release 9 to provide new data regarding the Hubble parameter $H(z)$.  From the sample of more than 130,000 massive and passively evolving galaxies, 5 new cosmology-independent $H(z)$ measurements have been obtained in the redshift range $0.3 < z < 0.5$, with an accuracy of $\sim11\%$ to $\sim 16\%$ incorporating both statistical and systematic uncertainties.
These new data were crucial to provide the first cosmology-independent determination of the transition redshift $z_t = 0.4\pm0.1$ between dark energy and dark matter dominated expansion. This result significantly disfavored the null hypothesis of no transition between decelerated and accelerated expansion at 99.9\% confidence level. This analysis highlighted the potential of cosmic chronometers to constrain the expansion history of the universe in a way competitive with standard probes.
The latest measurements of 31 Hubble parameters $H(z)$ from the galaxy differential age method, covering the redshift range $0.070 < z < 1.965$ are shown in Fig.3.
Actually, there is another approach to obtain the $H(z)$ data based on
the radial baryon acoustic oscillations (BAO) features from galaxy clustering \citep{Gaztaaga2009,Blake2012,Samushia2013,Font-Ribera2014,Delubac2015}.
However, the expansion rates obtained employing this method are dependent on an assumed fiducial cosmological
model and the prior for the distance to the last scattering surface from CMB observations \citep{Li2016c}, which is not quite suitable for our model-independent analysis. Consequently, we use only the latest 31 CC $H(z)$ measurements shown in Figure 3. Let us remind that $H(z)$ besides the $D_L(z)$ function, is the second ingredient necessary to implement Eq.(\ref{eq6v2}).

\section{Gravitational Waves from DECIGO as Alternative to Cosmic Chronometers}\label{subsec:GW}

Considering the redshift range of currently available cosmic chronometers data, it is tempting to seek for other complementary tools reaching higher redshifts. On the other hand, the era of gravitational wave (GW) astronomy just begun with the first detection of GW150914 signal and the total number of events registered so far $\sim 90$. Planned future detectors like the Einstein Telescope (ET) \footnote{The Einstein Telescope Project, https://www.et-gw.eu/et/.}, and satellite missions like LISA will provide orders of magnitude richer statistics of events probing the redshift range far deeper than optical surveys \citep{Amaro-Seoane2017}.
In particular, \citet{Kawamura2011,Seto2011} proposed a future
space gravitational wave detector called as DECi-hertz Interferometer Gravitational-wave Observatory (DECIGO). Its deci-Hertz frequency range fills the gap between LISA and Earth based interferometric detectors allowing to discover inspiralling binary system (BH-BH, BH-NS, NS-NS) up to a few years before they enter the frequency band of LIGO-Virgo-Kagra (or future ET) detectors \citep{Cao2022a,Cao2022b}.

DECIGO can become a unique tool to probe the cosmic expansion \citep{Schutz1986,Nishizawa2011} due to the following advantages compared with ground-based GW detectors: (i) lower-frequency band, (ii) larger number of GW cycles registered in a pre-merger phase,  (iii) longer observation time for each binary, and (iv) larger number of NS-NS binaries that can be detected up to $z\sim5$ \citep{Kawamura2019}. The accelerated expansion of the universe produce an additional phase shift in gravitational waveforms, which is analogous to the redshift drift in the electromagnetic domain \citep{Seto2001}. Focusing on the GWsignals from the binary system with component masses $m_1$ and $m_2$, the observed GW (time-domain) waveform can be written as $h(\Delta{t})$ and the Fourier transform of this waveform can be expressed as
\begin{equation} \label{Fourier}
\widetilde{h}(f)=\int_{-\infty}^{+\infty} dt e^{2\pi ift}h(\Delta{t})
\end{equation}
where $\Delta{t}\equiv{t_c-t}$ means the time to coalescence measured in the observer frame with $t_c$ representing the coalescence time.
If one relates the observed time interval $\Delta t$ with the respective time at the emitter frame $\Delta{t_e}$ taking into account both the time dilation in an expanding universe and the acceleration, the result is:
\begin{equation}
\Delta{t}=\Delta{T}+X(z_c)\Delta{T}^2,
\end{equation}
where $\Delta{T} = (1+z_c)\Delta{t_e}$, $z_c$ is the redshift of the source (coalescing binary) and the acceleration parameter $X(z)$ is related to the Hubble parameter $H(z)$ as follows
\begin{equation}
X(z)=\frac{1}{2(1+z)} \left[ H_0 (1+z) - H(z)\right].
\end{equation}

Then, $h(\Delta{t})$ can be re-expressed as a function of $\Delta T(\Delta t)$ with $h(\Delta{T})$ meaning the GW waveform without cosmic acceleration.
Substituting this to the Eq.(\ref{Fourier}) one can obtain
\begin{equation} \label{Fourier_accel}
\begin{split}
\widetilde{h}(f)=&e^{2\pi ift_c}\int_{-\infty}^{+\infty}d\Delta{T^{'}}e^{-2\pi if\Delta{T^{'}}}h(\Delta{T^{'}})\\
&\times{e^{-2\pi ifX(z_c)\Delta{T^{'}}^2}}.
\end{split}
\end{equation}
Considering the stationary phase approximation \citep{Cutler1994}, Eq.(\ref{Fourier_accel}) can be re-expressed as
\begin{equation}
\widetilde{h}(f)=e^{i\Psi_{acc}(f)}\widetilde{h}(f)|_{no accel},
\end{equation}
where
\begin{equation}
\begin{split}
\Psi_{acc}(f)\equiv &-2\pi fX(z_c)\Delta{T}(f)^{2}\\
&=-\Psi_N(f)\frac{25}{768}X(z_c)M_z (\pi M_zf)^{-8/3},
\end{split}
\end{equation}
where $M_z\equiv(1+z_c)\frac{(m_1m_2)^{3/5}}{(m_1+m_2)^{1/5}}$ is the redshifted chirp mass, and $\Psi_N(f)\equiv\frac{3}{128}(\pi  M_zf)^{-5/3}$.
Morever,
\begin{equation}
\widetilde{h}(f)|_{no accel}=e^{2\pi ift_c}\int_{-\infty}^{+\infty}d\Delta{T^{'}}e^{-2\pi if\Delta{T^{'}}}h(\Delta{T^{'}}),
\end{equation}
corresponds to the gravitational waveform in the Fourier domain without cosmic acceleration -- more details can be found in \citet{Yagi2012}.

We need to take the accuracies of binary parameters $\theta^i = (lnM_z,ln\hat{\eta},\hat{\beta},t_c,\phi_c,\theta_s,\phi_s,\theta_l,
\phi_l,D_L,X)$ into consideration, where $\hat{\eta} = m_1m_2/(m_1+m_2)^2$, $\hat{\beta}$ and $\phi_c$ are related to the spin-orbit coupling and the coalescence phase. The direction of the
sources are represented by $\theta_s$ and $\phi_s$.
And the direction of the orbital angular momentum can be described by
$\theta_l$ and $\phi_l$. In order to estimate the uncertainty of $\theta^i$, we employ Fisher analysis with inner product
\begin{equation}
\Gamma_{ij}=4Re\int_{f_{min}}^{f_{max}}df\frac{\partial_{i}\widetilde{h}^{*}(f)\partial_{j}\widetilde{h}(f)}{S_n(f)},
\end{equation}
where the noise spectrum $S_n(f)$ of the DECIGO can be found in \citet{Kawamura2006}. Following this procedure, the accuracies can be estimated as $\Delta\theta^i = \sqrt{\Gamma^{-1}_{ii}}$. By marginalizing the other parameters,
the accuracy of the acceleration parameter $X$ can be defined as $\sigma_X = \sqrt{8\Gamma^{-1}_{jj}}$. We show more details in \citet{Zhang2022}.

The classical redshift distribution of the GW sources observed on Earth will be used in our work \citep{Sathyaprakash2010} and the NSs coalescence rate can be approximated as \citet{Schneider2001,Cutler2009}. In our simulation, we assume equal mass NSs binary system with 1.4$M_{\odot}$ based on the flat $\Lambda CDM$ ($\Omega_m$=0.315 and $H_0$=67.4km/s/Mpc). Recent analysis of \citep{Kawamura2019} suggests that the space-based GW detector DECIGO can detect up to 10,000 GW events up to redshift $z\sim5$ in one year of operation. Thus, we simulate a mock data of 10,000 GW events to be used for the invariance of the speed of light analysis. The 10,000 simulated acceleration parameter $X(z)$ are shown in Figure 4.

We also show the redshift distribution of different measurements in Figure 5. Blue broken line means the redshift distribution of Hubble parameter $H(z)$ from the cosmic chronometers, it has a smaller redshift range and number than supernovae and quasars. Fortunately, future gravitational wave detector DECOGO can provide us with much higher redshifts and lots of measurements of acceleration parameters $X(z)$ (related to $H(z)$), which can help us study the invariance of the speed of light in more earlier Universe.

\begin{figure}
\begin{center}
\includegraphics[width=0.9\linewidth]{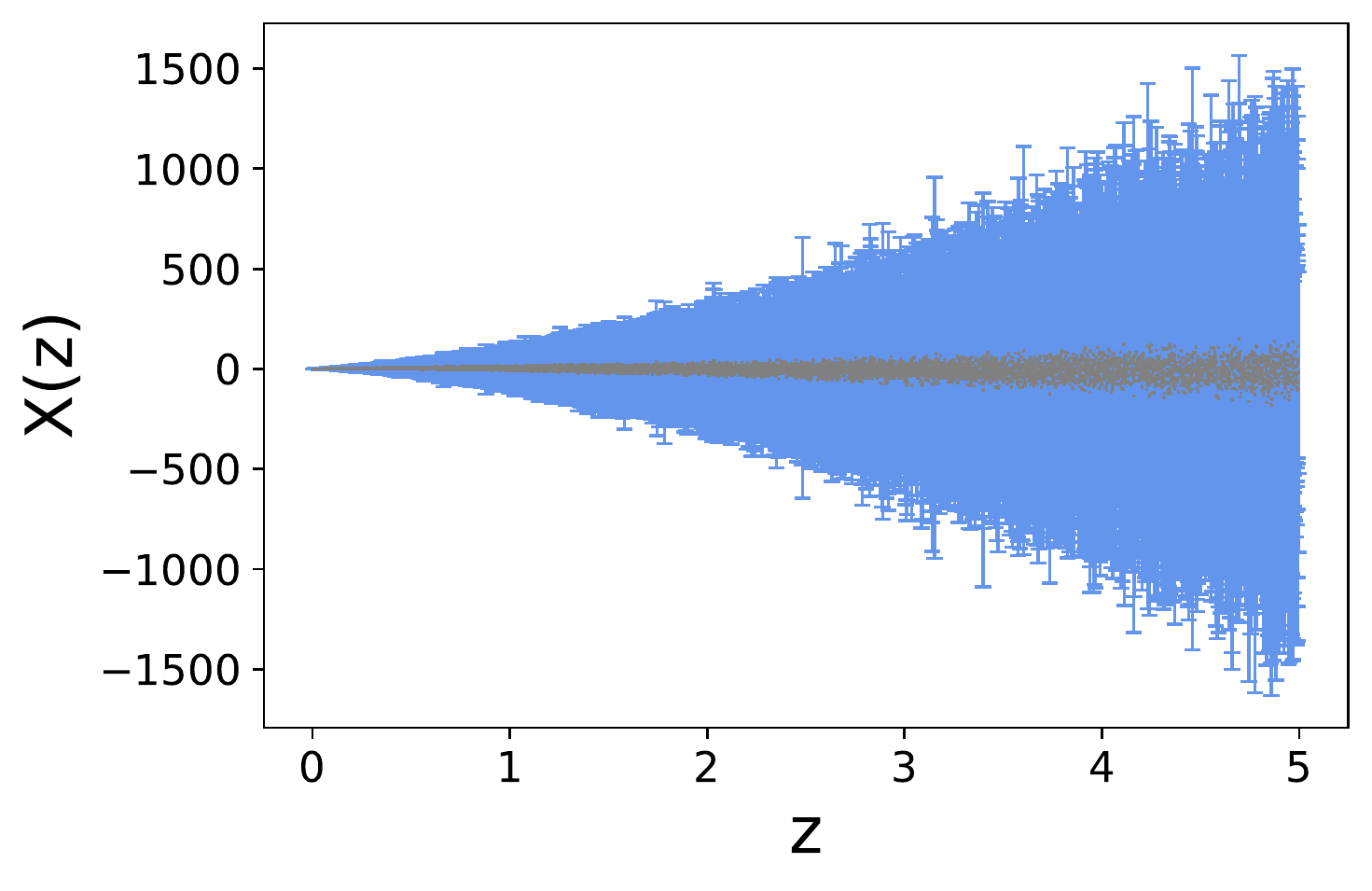}
\end{center}
\caption{Scatter plot of the 10,000 simulated acceleration parameter $X(z)$ based on the future space-based GW detector DECIGO. The gray point denotes acceleration parameter $X(z)$, with its $1\sigma$
uncertainty (blue bar).}
\end{figure}

\section{Results and Conclusions}\label{sec:result}

\begin{figure}
\begin{center}
\includegraphics[width=0.95\linewidth]{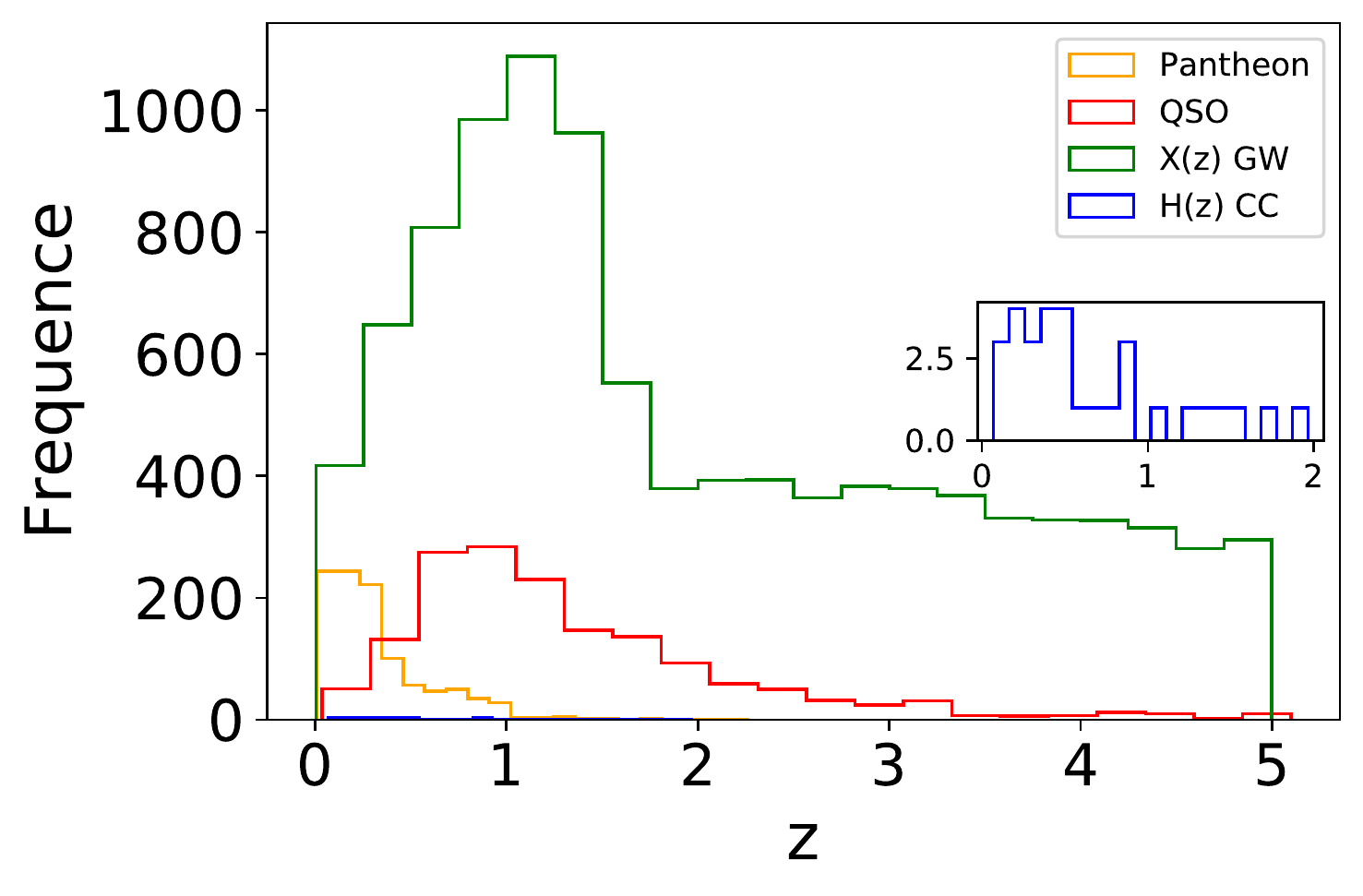}
\end{center}
\caption{Redshift distribution of different measurements used in this work. Blue, green, red, and orange lines correspond to redshift distributions of cosmic chronometers, GW sources used in simulating the acceleration parameters, UV-X QSO data and Pantheon SN Ia, respectively.}
\end{figure}

Applying methodology described above to determine the $\eta$ parameter measuring the constancy of $c$ from the combination of the Pantheon sample of \uppercase\expandafter{\romannumeral1}a supernovae reconstruction and 31 cosmic chronometers $H(z)$ measurements, we get the results shown in Figure 6.  The uncertainties have been calculated from the uncertainty propagation rule. We summarize the individual values of $\eta$ as the weighted mean with the inverse variance weighting \citep{Bevington1993}:
\begin{equation}
\eta=\frac{\Sigma_i\big(\eta_i/\sigma_{\eta_i}^2\big)}{\Sigma_i\big(1/\sigma_{\eta_i}^2\big)},\,\,\,\,\,\,\,\,\,
\sigma^2_{\eta}=\frac{1}{\Sigma_i\big(1/\sigma_{\eta_i}^2\big)},
\end{equation}
where $\eta$ stands for the weighted mean and $\sigma_{\eta}$ is its corresponding uncertainty. Following this most straightforward and popular way of summarizing multiple measurements, the result is Mean($\eta)=1.03\pm0.03$ with 31 measurements, which shows that SN Ia and $H(z)$ data do not indicate the deviation of $\eta$ from 1, i.e. there is no signal suggesting the different value of $c$ across the redshift range up to $z\sim 2$. In addition, we also summarized our findings with a robust non-parametric statistics by calculating the median and the corresponding median absolute deviation \citep{Zheng2016,Liu2023}. Considering that the probability that $n$-th observation is higher than the median follows the binomial distribution \citep{Feigelson2012}:
\begin{equation}
P=2^{-N}N!/[n!(N-n)!
\end{equation}
where $N$ is the total number of multiple measurements, one can also define the 68.3\% confidence interval with median statistics. Our assessment is Mean($\eta)=1.04\pm0.05$ with the median value and the absolute deviation. It should be noted that the above estimates are based on the prior of the Hubble constant $H_0=67.4\pm0.5$ km/s/Mpc from the latest \textit{Planck} CMB observations \citep{Aghanim2020}. Considering the redshifts of the Pantheon sample of \uppercase\expandafter{\romannumeral1}a supernovae are $z \sim2$ which correspond to the late universe, it is necessary to discuss the performance of the speed of light under different priors of Hubble constant. Therefore, we also use $H_0=74.03\pm1.42$ km/s/Mpc from the SH0ES collaboration \citep{Riess2019} to estimate the invariance of the speed of light. In this case, we obtain Mean($\eta)=0.93\pm0.02$ and Median($\eta)=0.94\pm0.05$ in the framework of weighted mean and median statistics, which produces a possible deviation from the constant speed of light up to $z\sim 2$. However, our results are still marginally consistent with $\eta=1$ within 2$\sigma$ confidence level, which is in full agreement with other recent tests involving cosmological data.

\begin{figure}
\begin{center}
\includegraphics[width=0.95\linewidth]{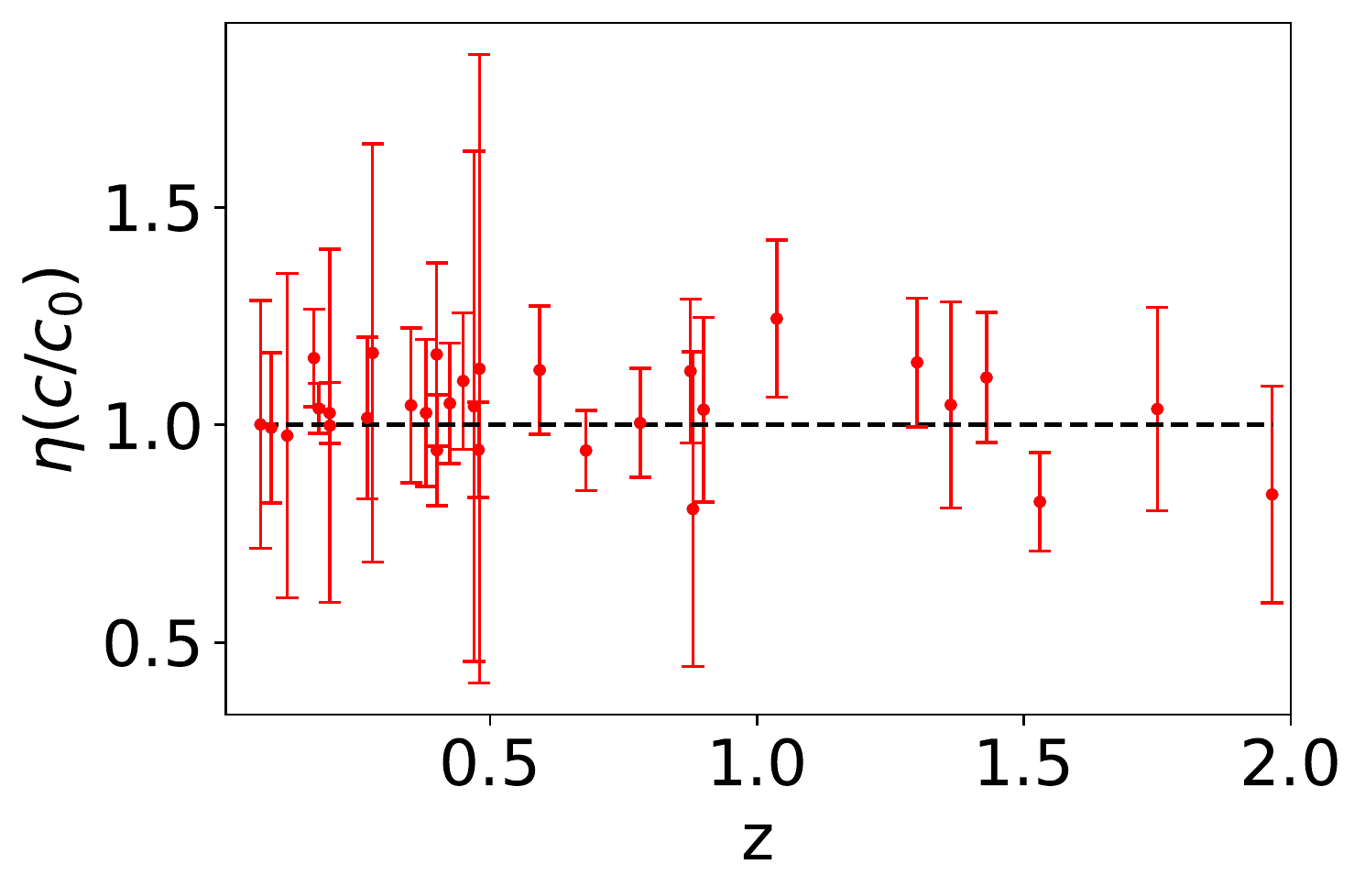}
\end{center}
\caption{Reconstruction of $\eta\equiv c/c_0$ parameter measuring the invariance of the speed of light from the Pantheon sample of \uppercase\expandafter{\romannumeral1}a supernovae and cosmic chronometers $H(z)$. Red dots represent 31 measurements of $\eta$ and their 1$\sigma$ uncertainties, and the black line shows the value  indicating the constancy of the speed of light.}
\end{figure}

The use of UV-X relation in high-redshift quasars instead of SN Ia was motivated by reaching farther in redshift with the assessment of $\eta$. In fact, because both $D_L(z)$ and $H(z)$ data are needed, we are able to infer $\eta$ parameter also up to $z\sim 2$. However, comparison between Pantheon + H(z) vs. UV-X QSO + H(z) allows to address the utility of UV-X QSO in the future.
As one can see in Figure 7, almost all of the reconstructed $\eta$ is consistent with the constancy of the speed of light within the $1\sigma$ confidence level. As for the summary statistics, the weighted mean value and corresponding uncertainty is Mean($\eta)=1.19\pm0.07$, while the median statistics turns out to be Median($\eta)=1.22^{+0.05}_{-0.13}$. We summarize our findings from $QSO+H(z)$ and $Pantheon+H(z)$ in Table 1. Compared with what is obtained from the \uppercase\expandafter{\romannumeral1}a supernovae, some $\eta$ measurements have demonstrated mild deviation from the standard case ($\eta=1$) within
the observational uncertainty, especially in the low-redshift range ($z<0.5$). Such possible tension between UV-X QSO and standard cosmological scenario has been recently traced and extensively discussed in \citet{Lian2021,Zheng2022}. Moreover, there is no obvious improvement in the precision when the UV-X QSO data are used. Most likely it is due to still significant intrinsic dispersion in the UV-X QSO data. However, the uncertainty bars of individual assessments of $\eta$ parameter support the future potential of UV-X QSO data in similar projects. The advantage of higher redshift coverage could not be taken, for the reasons already discussed. This possibility will be tested on the simulated data on acceleration parameter $X(z)$ from DECIGO. Before we go to this point, let us compare our findings with the earlier studies done using alternative probes. The precision of our inference regarding $\eta$ parameter is comparable to that attained in the study of the invariance of the speed of light from the recently compiled set of strong gravitational lensing (SGL) systems \citep{Liu2021}. Our conclusions that there is no clear evidence for the different value of the speed of light in earlier epochs is also consistent with the analysis of \citet{Cai2016,Cao2017}.

\begin{figure}
\begin{center}
\includegraphics[width=0.95\linewidth]{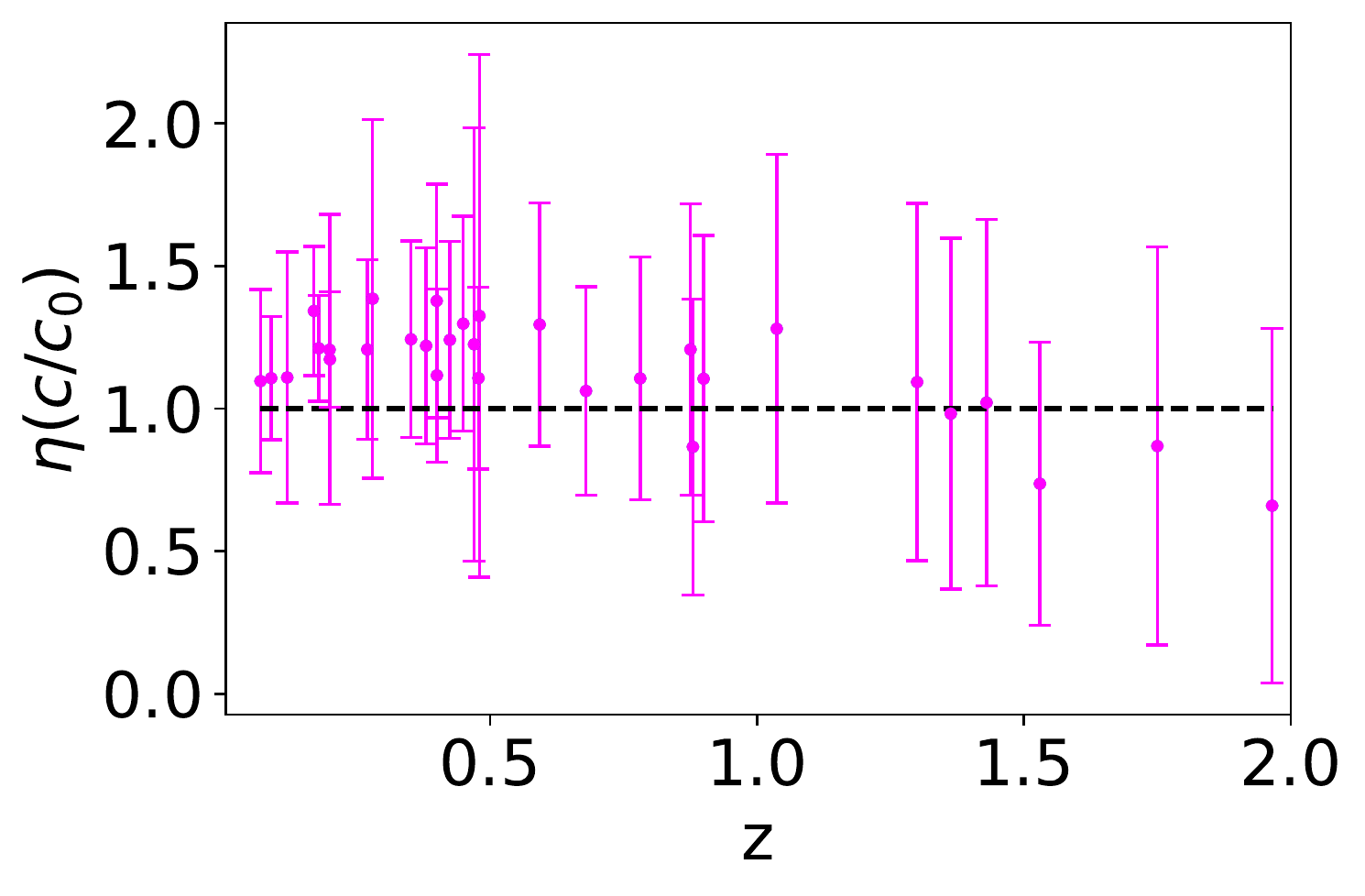}
\end{center}
\caption{Reconstruction of $\eta\equiv c/c_0$ parameter measuring the invariance of the speed of light from the UV-X relation in high-redshift quasars and cosmic chronometers $H(z)$. Pink dots represent 31 measurements of $\eta$ and their 1$\sigma$ uncertainties, and the black line shows the value  indicating the constancy of the speed of light.}
\end{figure}

\begin{table}
\begin{center}
\caption{The weighted mean and median values of $\eta$ parameter measuring the invariance of the speed of light from the $QSO+H(z)$ and $Pantheon+H(z)$ samples. The associated dispersion measures are: weighted standard deviation and median absolute deviation.}

\begin{tabular}{c c c } \hline\hline
       & Mean($\eta)$  & Median($\eta)$  \\
\hline
$Pantheon+H(z)$    & $1.03\pm0.03$    & $1.04\pm0.05$  \\
 $QSO+H(z)$    &  $1.19\pm0.07$     & $1.22^{+0.05}_{-0.13}$ \\
\hline
\end{tabular}

\end{center}
\end{table}

Contemplating alternative measures of the expansion rate $H(z)$ covering higher redshifts, we turned to the future space-borne GW detector DECIGO. Using the simulated data of 10,000 the acceleration parameter $X(z)$ measurements from DECIGO, combined with the UV-X QSO Hubble diagram, we obtained $\eta=1.016\pm0.002$. Considering instead, the Pantheon sample, the $\eta$ parameter measuring the invariance of the speed of light can be constrained as $\eta = 1.002\pm0.001$. We summarize our results in Table 2. They are also illustrated in Fig. 8.
The expected precision is promising and is comparable to the forecasts of \citet{Cao2020}, who predicted that multiple measurements of galactic-scale strong gravitational lensing systems with Type Ia supernovae acting as background sources from the future Legacy Survey of Space and Time  (LSST) would be able to constrain $\Delta{c}/c$ at the level of $10^{-3}$. One should also note that this last approach we discussed was an attempt to measure the speed of light by combining the GW data with electromagnetic ones, which would also have a potential  for testing general relativity.
\begin{figure}
\begin{center}
\includegraphics[width=0.9\linewidth]{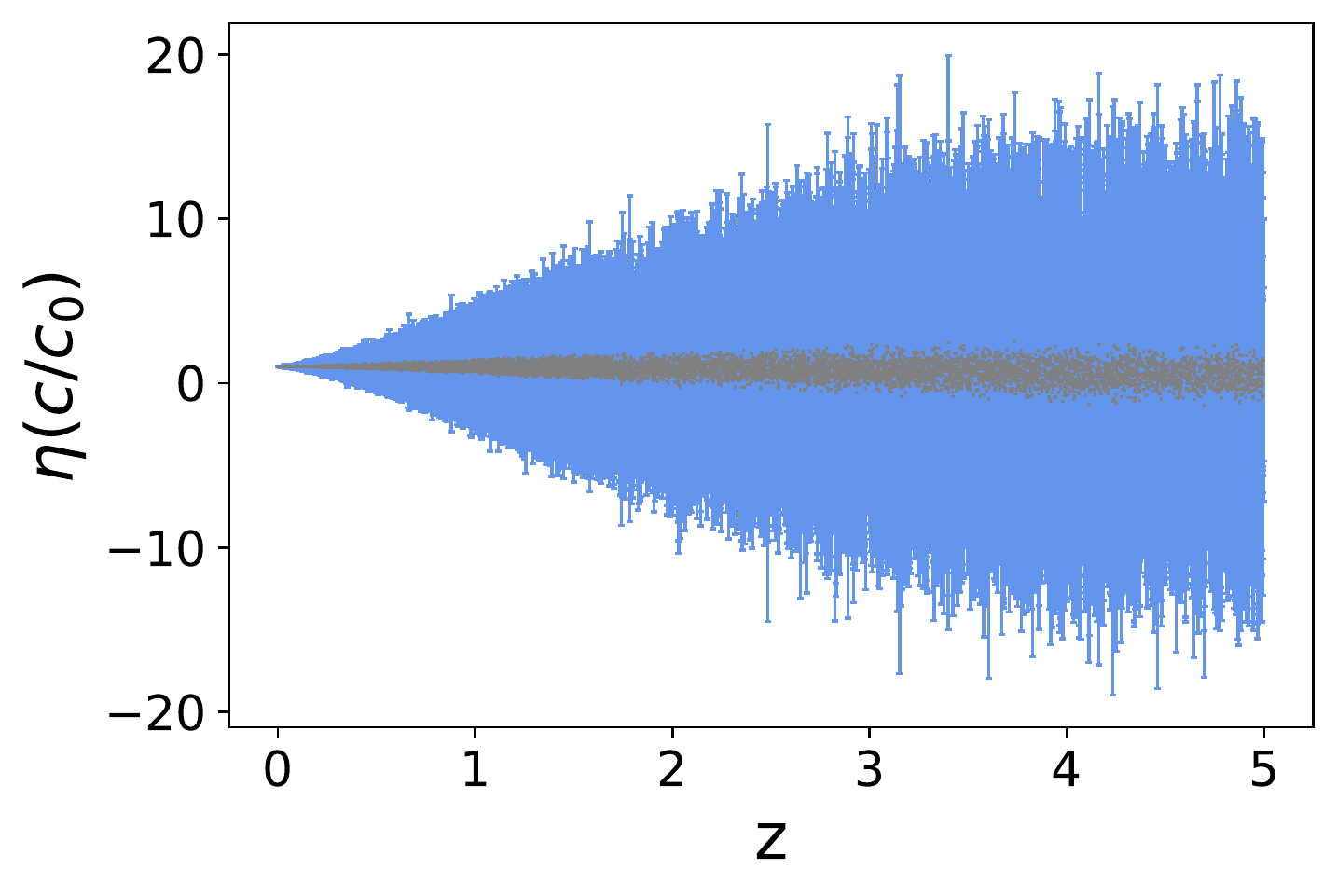}
\includegraphics[width=0.9\linewidth]{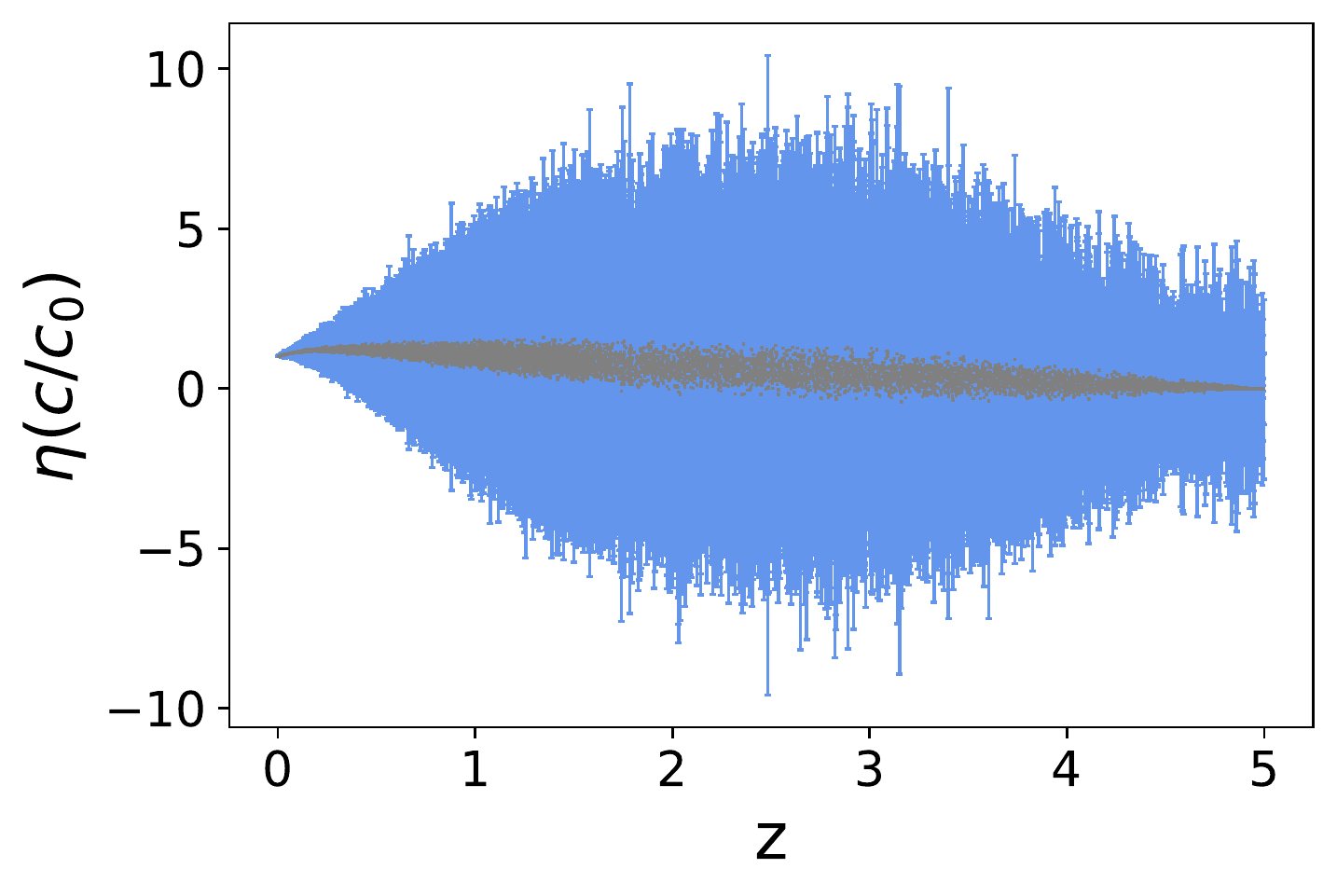}
\end{center}
\caption{Reconstruction of the invariance of the speed of light $\eta\equiv c/c_0$ from the Hubble diagram of  type \uppercase\expandafter{\romannumeral1}a supernovae (upper), high-redshift (lower) quasars and the simulated acceleration parameter $X(z)$ based on the standard siren of GW space-based detector DECIGO. The gray dots with blue bars represent 10,000 measurements of $\eta$ and their 1$\sigma$ confidence level.}
\end{figure}

\begin{table}
\begin{center}
\caption{$\eta$ parameter measuring the invariance of the speed of light from the $QSO+X(z)$ and $Pantheon+X(z)$ samples and its precision.}
\begin{tabular}{c c c } \hline\hline
       &$\eta$  &  $\Delta\eta$ \\
\hline
$Pantheon+X(z)$    & $1.002\pm0.001$    & $10^{-3}$  \\
 $QSO+X(z)$    &  $1.016\pm0.002$     & $10^{-3}$ \\
\hline
\end{tabular}

\end{center}
\end{table}

In conclusion, we proposed an original technique to test the invariance of the speed of light with the Hubble diagrams of standard candles (SN Ia up to $z\sim2.4$ and UV-X QSO relation up to $z\sim5.1$) combined with Hubble parameters $H(z)$ inferred from cosmic chronometers. Moreover, this method does not rely on the details of the cosmological model: only flat FLRW metric is assumed. Generalization to non-flat FLRW metric is straightforward. Moreover, precise and accurate measurements of $H_0$ and curvature parameter, which would be the input to our method are anyway more than welcome in any cosmological studies. Our findings confirm that there is no clear evidence for the deviation of $\eta$ from 1 at the current observational data level. This result  supports the claim that the speed of light is a fundamental constant of nature. It should be noted that due to the restricted redshift range and small number of cosmic chronometers, we were not able to take the full advantage of the high redshift reach of standard candles (UV-X QSO in particular). However, the results obtained with simulated acceleration parameter $X(z)$ expected from the future DECIGO mission are encouraging: the attainable precision of testing the invariance of the speed of light is at the level of $10^{-3}$. The combination of measurements obtained form GW and electromagnetic windows additionally opens a possibility to test the general relativity in a broader sense.

\acknowledgments

This work was supported by the National Natural Science Foundation of China under Grants Nos. 12021003, 11690023, and 11920101003; the Strategic Priority Research Program of the Chinese Academy of Sciences, Grant No. XDB23000000; and the Interdiscipline Research Funds of Beijing Normal University. Y.T. Liu was supported by the Interdiscipline Research Funds of
Beijing Normal University (Grant No. BNUXKJC2017) and China Scholarship Council (Grant No. 202106040084). M. Biesiada was supported by Foreign Talent Introducing Project and Special Fund Support of Foreign Knowledge Introducing Project in China (No. G2021111001L).

\end{document}